\documentclass[twocolumn,showpacs,amsmath,amssymb,nofootinbib]{revtex4}
\usepackage{graphicx}
\usepackage{dcolumn}%
\usepackage{bm}
\usepackage{setspace}

\begin{document}
\title{Double strand breaks in DNA resulting from double-electron-emission events}
\author{Eugene Surdutovich$^{1,2}$ and Andrey V. Solov'yov$^{2}$\footnote{On
leave from A.F. Ioffe Physical Technical Institute, St. Petersburg,
Russia} } \affiliation{ $^1$Department
of Physics, Oakland University, Rochester, Michigan 48309, USA\\
$^2$Frankfurt Institute for Advanced Studies,
Ruth-Moufang-Str. 1, 60438 Frankfurt am Main, Germany\\
}
\date{\today}
\begin{abstract}
A mechanism of double strand breaking (DSB) in DNA due to the action
of two electrons is considered. These are the electrons produced in
the vicinity of DNA molecules due to ionization of water molecules
with a consecutive emission of two electrons, making such a
mechanism possible.  This effect qualitatively solves a puzzle of
large yields of DSBs following irradiation of DNA molecules. The
transport of secondary electrons, including the additional
electrons, is studied in relation to the assessment of radiation
damage due to incident ions. This work is a stage in the inclusion
of Auger mechanism and like effects into the multiscale approach to
ion-beam cancer therapy.
\end{abstract}

\pacs{87.53.-j, 61.80.-x, 34.50.Gb, 36.40.Cg}

\maketitle

The analysis and assessment of radiation damage are important in
relation to a wide range of applications from cancer therapy to
radiation protection of humans and electronics. Studies of biodamage
due to irradiation with ion beams, centered on the analysis of
pathways of DNA damage, are discussed in a large number of papers
observed in reviews~\cite{FokasKraft09,SchardtRMP10}; the multiscale
approach to the physics of ion-beam cancer
therapy~\cite{pre,precomplex} is also among these studies.
A double strand break (DSB) in DNA is the most pernicious kind of
damage that happens as a result of radiation impact. It is defined
as the simultaneous breaking of the two strands of a DNA molecule
within the distance of 10~base pairs along the helix, which
corresponds to a single convolution of a DNA molecule. This type of
lesion is emphasized because of the difficulties of its repair and
thus close connection to cell lethality. What processes lead to this
lesion?

If the primary ionizing projectiles are ions, a substantial fraction
of damage is done by secondary electrons, formed in the process of
ionization of the medium. A number of different pathways of damage
due to these electrons were considered in
Refs.~\cite{precomplex,Kiefer,Sevilla11}. The direct measurements of
DNA damage due to incident electrons were presented in, e.g.,
Ref.~\cite{DNA3}, which brought to light the possibility that low
energy electrons were important agents of DNA damage.
Since then, the mechanism of a SSB due to the action of a single
electron, related to the formation of a transient negative ion (TNI)
as a part of the process of dissociative electron attachment (DEA)
has been widely discussed in the literature, e.g.,
Refs.~\cite{Gianturco1,Sanche11, Sevilla11} with emphasis on low
(under ionization threshold) energy of the incident electrons.

Unexpectedly high yields of DSBs compared with SSBs in
Ref.~\cite{DNA3} lead to a hypothesis that DSBs can be caused by a
single electron. This logic has been widely accepted and used by the
present authors to calculate the yields of DSBs due to
ions~\cite{pre}. However, the mechanism of DSBs due to low-energy
electrons is still quantitatively unclear despite qualitative
arguments, suggesting that the breaks in the second strand are due
to the action of debris generated by the first SSB~\cite{DNA3}. In
Ref.~\cite{pre}, the production of DSBs by two separate electrons
was also considered, but that analysis was then shelved, since the
number density of secondary electrons due to primary ionization with
an ion was not nearly enough for this effect to be considerable in
comparison with DSBs due to single electron action.

In this work, we argue that additional electrons emitted in the
vicinity of a DNA molecule as a result of the Auger-like mechanism
augment the above effect and thus constitute a mechanism for a DSB.
We explore the probability of two electrons, produced in the
vicinity of a DNA molecule, be incident on a single convolution of
this molecule. These additional electrons emerge as a result of
double ionization events such as the Auger effect in single
molecules, e.g., Refs.~\cite{Aug1,Aug2,Aug3,Aug4} for water, or to
the effect of intermolecular coulombic decay (ICD), studied in
Refs.~\cite{cederbaum97,cederbaum06,cederbaum11,Doerner, Becker} for
water clusters. The actual mechanism of double ionization is not
important for our analysis. We will use a generic term
``double-ionization-events'' to describe all relevant events leading
to a production of additional electrons, and we will refer to the
additional electron as an Auger electron, even if it originates from
ICD.

\begin{figure}
\includegraphics[width=2.0in]{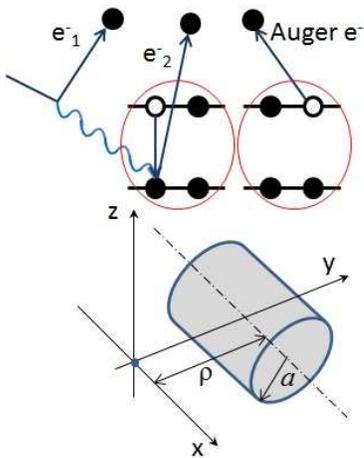}
\caption{The scheme of ionization with the Auger mechanism applied
to ICD; $e_2$ and the Auger electron emerge from different
molecules. The geometry of the problem; ionization happens at the
origin; the cylinder represents a DNA convolution.}
\label{fig.geom}       
\end{figure}

\noindent{\bf 1.} Auger electrons are produced as a result of
non-radiative relaxation of holes produced by primary and secondary
ionization. 
They may emerge consequent to the ionization of water or other
molecules or clusters of the medium.
 If a secondary
electron ionizes a molecule by kicking out one of its inner-shell
electrons, a hole is formed on this molecule. If this hole then
relaxes via a non-radiative channel (an Auger electron is emitted
from the same molecule or an adjacent molecule in a cluster) the
total number of electrons, emerging from a sub-nm locality, is equal
to {\em three}, the ionizing electron, the released electron, and
the Auger electron, as shown in Fig.~\ref{fig.geom}.
If a double-ionization event occurs on a DNA molecule, the
probability of damage, such as a DSB, comprises the sequence of
these processes with a corresponding dynamics of the DNA molecule
and possible further interaction with these three electrons. If a
water molecule (cluster) of the medium located in the vicinity of a
DNA molecule is a host of a double-ionization event, then these
three electrons can independently diffuse and stumble onto a DNA
convolution and produce two SSBs, possibly yielding a DSB. Let us
consider the latter scenario in more detail.

It is interesting to notice that the energies of each of the three
electrons, emerging from a water molecule or cluster, are likely to
be below 15~eV. An average ionizing electron has the energy around
45-50~eV, corresponding to the average energy of electrons produced
by an ion~\cite{epjd}. After this ionization event, it loses about
30~eV~\cite{cederbaum06}.\footnote{In order to kick out an electron
from $1a_1$ MO of ${\rm H_2O}$ molecule, the secondary electron has
to have the energy above 540~eV. Such electrons ere very rare.
Ionization of other state cannot produce Auger electron because of
the lack of energy.} Thus, the ionizing electron is left with less
than 15~eV while the other two will have even smaller
energies~\cite{cederbaum11}. These low-energy electrons are likely
to be engaged into the DEA channel leading to SSBs. How does this
scenario make a difference in the DSBs production?

 The number density of secondary electrons, produced on the ion's path,
reduces rather steeply with the increasing distance from the path,
so that the probability of two electrons incident on a single
convolution of a DNA molecule, located a few nm from the path is
very small. However, when ionization of a water molecule (cluster)
producing an Auger electron occurs at a distance from the path,
three electrons emerging from the same spot substantially boost the
local number density of electrons and, hence, the probability of a
nearby DNA convolution to be hit with two electrons. A similar
process may take place if the primary projectile is a photon. In
that case, two
(instead of three) electrons are produced in a locality and are
capable of producing a DSB in a DNA molecule. If incident photons
ionize the $1a_1$ state, then a cascade of Auger electrons may
follow, rapidly increasing the number density of electrons.

In order to give a quantitative example of the effect caused by the
Auger mechanism, let us consider an event of ionization with Auger
emission caused by a secondary electron, produced by an incident
carbon ion. The goal of this example is to calculate the transport
of three electrons emerging from the ionization locality, which we
will treat as a point, to a nearby DNA convolution.

Let an event of double ionization of a water molecule (cluster)
happen at the origin. We consider a three-dimensional random walk of
electrons from this point. We represent a single DNA convolution
with a cylinder of radius $a=1.15$~nm and a length of 3.4~nm. For
simplicity and definiteness, we situate this cylinder symmetrically
with respect to the $y$-axis, with the cylinder's axis parallel to
the $x$-axis and lying in the $xy$ plane, as shown in
Fig.~\ref{fig.geom}. The distance between the origin and the
cylinder's axis, $\rho$, exceeds $a$ so that the Auger electron is
emitted outside the convolution.

If one electron is emitted from the origin at $t=0$, according to
Ref.~\cite{Chandra}, its rate, $dp_1/dt$, of passing through the
patch, $d{\vec A}$, located at a distance $r$ from the origin, is
given by the expression
\begin{eqnarray}
\frac{dp_1(\vec r, t)}{dt} = d{\vec A}\cdot D {\bf
n_r}\frac{\partial P(t, r)}{\partial
  r}~, \label{mult1t}
\end{eqnarray}
where $D={\bar v} l/6$ is the diffusion coefficient, ${\bar v}$ is
the speed of the electron, ${\bf n_r}$ is a unit vector in the
radial direction, and
\begin{eqnarray}
P(t, r)=\left(\frac{3}{2 \pi {\bar v}t
l}\right)^{3/2}\exp\left(-\frac{3r^2}{2 {\bar v}t l}\right)
\label{rwalk2t}
\end{eqnarray}
is the probability density to observe a randomly walking electron at
a time $t$ and a distance $r$ from the origin. Eq.~(\ref{mult1t})
should be integrated over both the time and $d{\vec A}$, in order to
calculate the probability for the electron to encounter the
cylinder.

The time dependence in Eqs.~(\ref{mult1t}, \ref{rwalk2t}) can be
translated into the dependence on number of steps, $k$, using ${\bar
v}t=kl$, where $l$ is the elastic mean free path of electrons in the
medium. Then, we rewrite Eq.~(\ref{mult1t}), substituting
(\ref{rwalk2t}), and switching from variable $t$ to $k$ as
\begin{eqnarray}
dp_1(\vec r, k)= dk d{\vec A}\cdot {\bf n_r}
    \frac{r}{2 k} \left(\frac{3}{2\pi k l^2}\right)^{3/2}
\exp\left(-\frac{3 r^2}{2 k l^2}\right) \label{mult4}
\end{eqnarray}
and integrate it over $k$ (from the minimal number of steps
necessary to reach the surface of the cylinder to infinity) and $dA$
over the surface of the cylinder. As was noticed above, these
electrons are very near or below the ionization threshold and their
interactions with water molecules will be, by and large, elastic.
Therefore, we integrate Eq.~\ref{mult4} without including
attenuation due to inelastic collisions and using the value for the
mean free path, $l=0.15$~nm~\cite{Tung}.

\begin{figure}
\includegraphics[width=2.7in]{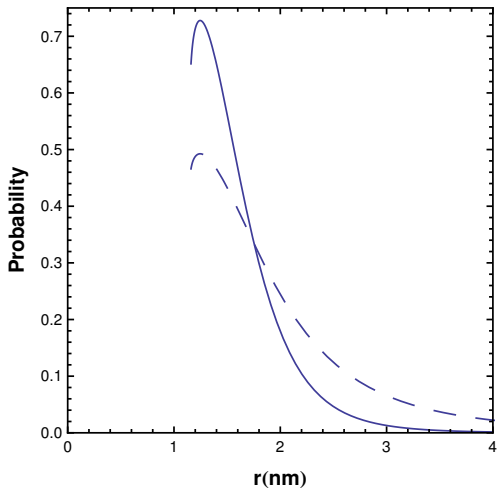}
\caption{The probability for two electrons to pass through a single
convolution of DNA (solid line) compared to that of one electron
(dashed line).}
\label{fig.flux3}       
\end{figure}

The results of this integration of are presented in
Fig.~\ref{fig.flux3}. The probability decreases with increasing $r$.
This probability is proportional to that of producing a strand break
or a different type of lesion on the length of a given DNA
convolution. The corresponding coefficient is the probability of
inducing an SSB in a single electron action, $\Gamma$, remains
largely unknown and its calculations and measurements remain a task
for atomic physicists and quantum chemists.

The second and third electrons from the three, produced in the
double ionization event, are independent from the first one and,
hence, the probability of two electrons to impact the cylinder,
$p_2$, is given by
\begin{eqnarray}
p_2(r)=3 p_1^2(r)~. \label{prob-pass2}
\end{eqnarray}
The dependence of this probability on the distance of the cylinder
axis from the origin is shown in Fig.~\ref{fig.flux3}.
It is remarkable, that for $r<2$nm the probability of the impact of
two electrons is comparable to that of one. Of course,
Fig.~\ref{fig.flux3} includes neither the value of $\Gamma$, nor the
probability of an ionization event with the following Auger
emission, but still, a substantial quantity of this fluence makes
the Auger-mechanism influence on DSB yield viable.

\noindent{\bf 2.} Auger electrons emitted from the ion's path as a
result of primary ionization propagate similarly to secondary
electrons, so the correction to the number of agents interacting
with a DNA molecule situated at a distance from the path can be made
after the calculation of the probability of Auger electron emission
during the primary ionization. Our goal in this section is to
consider the Auger electrons emitted as a result of the ionization
of water molecule clusters by secondary electrons.

\begin{figure}
\includegraphics[width=3.0in]{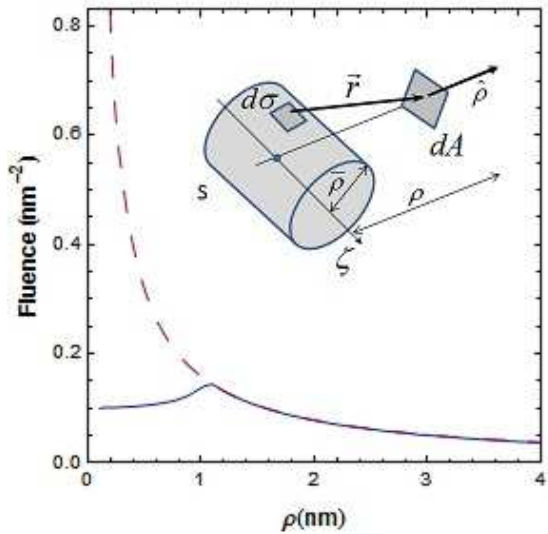}
\caption{The dependence of the fluence of an Auger electron from
ionization by a secondary electron on the distance from the path
(solid line) compared to that of a single electron emitted from the
path (dashed line).}
\label{fig.cylAuger}       
\end{figure}

The scenario is as follows: Secondary electrons originate on the
ion's path and diffuse (through a random walk) away from the path.
Their interactions with water molecules are primarily elastic until
they ionize a water molecule (cluster). This happens about once per
every 30 elastic collisions at relevant energies of secondary
electrons. A typical secondary electron ionizes one water molecule
before it becomes thermalized and finally bound. More rare energetic
$\delta$-electrons are not considered in this discussion. For
simplicity, let us assume that ionization happens on a certain step
in the random walk of a 45-eV electron. At that time, this electron
is most likely situated at a distance from the path, given by the
expression,
\begin{eqnarray}
{\bar \rho}=\int \rho P(k, r)d^3r~, \label{raverage}
\end{eqnarray}
where $k=30$ and $\rho$ is the radial distance from the path. The
calculations, with the above parameters for secondary electron
propagation in water, give ${\bar \rho}$ approximately equal to
1~nm. This means that the locus of first ionization events is a
cylinder with a radius of 1~nm on the axis of the ion path. Then, we
can calculate the fluence at a distance $\rho$ from the path due to
Auger electrons emitted from the surface of this cylinder. It is
given by the integration of Eq.~(\ref{mult4}) normalized  by $dA$
over the surface of the cylinder:
\begin{eqnarray}
p_A(\rho)=\int_S\left|\frac{\vec r}{r}\cdot{\bf n_\rho}\right|\frac
{dp_1(r, k)}{dA}\frac{d\sigma}{S} \label{probfromcyl}
\end{eqnarray}
where $r^2={\bar \rho}^2+\zeta^2+\rho^2-2{\bar \rho}\rho \cos \phi$,
$\zeta$ is the coordinate along the path, $\phi$ is the azimuthal
angle, and $\bf n_\rho$ is a unit vector in the radial direction.
The rest of the geometry is shown in Fig.~\ref{fig.cylAuger} along
with the results for $p_A(\rho)$. The fluence per one electron from
the cylinder is compared to that of electron coming from the path.
Starting from a small distance from the cylinder, the fluences are
equal, which means that outside some domain, the total fluence is
equal to the fluence of secondary electrons emitted on the path
multiplied by $(1+2\psi)$, where $\psi$ is the probability of the
double-ionization event on impact and the factor of two is due to an
extra electron emitted in this process.

\noindent{\bf 3.} The quintessence of this paper is that due to the
Auger mechanism, events in which two or more electrons interact with
a single DNA convolution are not rare. Double strand breaks or other
types of complex damage may result from these interactions. Thus,
this effect qualitatively provides  a solution to the puzzle of high
yields of DSBs by finding sources of high local electron density
near sites of damage. Since the Auger emission may result from the
ionization of a water molecule by an incident photon, the above
two-electron mechanism can also contribute to the DSB yield
consequent to an irradiation of tissue with photons.

We considered the major concepts related to the transport of Auger
electrons and made first steps to including the Auger electrons in
the multiscale approach to the calculation of radiation damage by
ions. Besides the consideration of a two-electron mechanism for
DSBs, our findings also correct the fluence of secondary electrons
through the DNA convolution due to the Auger mechanism. The
inclusion of this effect into the multiscale approach is an
important step, since, in general, the Auger mechanism plays a
significant role in radiation damage~\cite{Prise11}.

Future research in this route will be based on the calculations and
measurements of the cross sections of ionization of water molecules
and clusters with a consequent emission of Auger electrons and the
cross sections of DSBs due to two incident electrons.

The calculation of the fluence of Auger electrons presented in the
previous section gives a framework for the calculation of the
transport of free radicals formed due to an ion's traverse through a
medium. This calculation, without the use of Monte Carlo simulations
has long been desired. Finally, the analysis presented in this paper
can be extended to the calculation of complex damage as in
Ref.~\cite{precomplex} with the inclusion of the Auger mechanism.

\begin{acknowledgments}
ES is grateful to J.S. Payson who critically read the manuscript and
the support of COST Action MP1002 ``Nano-scale insights in ion beam
cancer therapy.''
\end{acknowledgments}


\end{document}